# Quasi-phasematching of harmonic generation via multimode beating in waveguides

**B. Dromey, M. Zepf**

*Plasma and Laser Interaction Physics, Department of Physics and Astronomy Queens University Belfast, United Kingdom, BT7 1NN.*
b.dromey@qub.ac.uk, m.zepf@qub.ac.uk

**M. Landreman, S. M. Hooker**

*Department of Physics, University of Oxford, Clarendon Laboratory, Oxford, Parks Road, Oxford, OX1 3PU*

**Abstract:** A new scheme for quasi-phasematching high harmonic generation (HHG) in gases is proposed. In this, the rapid variation of the axial intensity resulting from excitation of more than one mode of a waveguide is used to achieve quasi phasematching. Numerical modeling demonstrates enhancement of the harmonic signal over that achieved for a single coherence length by factors $>10^4$.

©2007 Optical Society of America

**OCIS codes**: (190.4160) Multiharmonic generation

## References and links

1. A. L'Huillier, P. Balcou, and L. Lompre, "Coherence and Resonance Effects in High-Order Harmonic Generation," Phys. Rev. Lett. **68**, 166-169 (1992). P. B. Corkum "Plasma perspective on strong field multiphoton ionization" Phys. Rev. Lett. **71,** 1994 (1993).
2. C. Spielman, *et al*., "Generation of Coherent X-ray Pulses in the Water Window Using 5 fs Laser Pulses," Science **278**, 661-664 (1997).
3. X. F. Li, A. L'Huillier, M. Ferray, L. A. Lompr´e and G. Mainfray, Multiple-harmonic generation in rare gases at high laser intensity, *Phys. Rev. A* **39**, 5751 (1989).
4. C. Durfee, A. Rundquist, S. Backus, C. Herne, M. M. Murnane, and H. C. Kapteyn, "Phase Matching of High-Order Harmonics in Hollow Waveguides," Phys. Rev. Lett. **83**, 2187-2190 (1999).
5. A. Paul, *et al*., Quasi-phase-matched generation of coherent extreme ultraviolet light, *Nature*, 42, 51–54, (2003).
6. I. P. Christov, H. C. Kapteyn and M. M. Murnane, "Dispersion-Controlled Hollow Core Fiber for Phase Matched Harmonic Generation" Opt. Express **3**, 360 (1998).
7. S. Voronov "Control of Laser High-Harmonic Generation with Counterpropagating Light" Phys. Rev. Lett. , **87**, 133902 (2001).
8. E. A. Gibson, *et al* Coherent soft x-ray generation in the water window with quasi-phase matching. *Science* **302**, 95–98, (2003).
9. R. L. Abrams, "Coupling losses in hollow waveguide fiber resonators" IEEE J. Quantum Electron. **8**, 838 (1972).
10. C. Courtois, A. Couairon, B. Cros, J. R. Marques G. Matthieussent, Propagation of in tense ultrashort laser pulses in a plasma filled capillary tube: Simulations and experiments, *Physics of Plasmas* **8**, 3445, (2001).
11. Marcatili, E. A. J. and R. A. Schmeltzer. Hollow metallic and dielectric waveguides for long distance optical transmission and lasers. Bell Syst. Tech. J., **43**, 1783 (1964).
12. Cros, B. et al. Eigenmodes for capillary tubes with dielectric walls and ultraintense laser pulse guiding. Phys. Rev. E, **65**, 026405 (2002).
13. M. Ammosov, N. Delone, and V. Krainov, "Tunnel ionization of complex atoms and of atomic ions in an alternating electromagnetic field" , *Sov. Phys. JETP* **64**, 1191 (1986).
14. M. Zepf and B. Dromey, Queens University Belfast, are preparing a manuscript to be called "Bright quasi-phase matched soft x-ray harmonic radiation from Argon ions" (Feb 2007) -

## 1. Introduction

Efficient generation of a coherent beam of femtosecond pulses at soft x-ray wavelengths may have applications such as high temporal and spatial resolution imaging of hydrated biological samples, photoemission spectroscopy and metrology of silicon wafers for next generation lithography. One approach for generating such beams is high harmonic generation (HHG) in gaseous targets [1, 2, 3]. To date, the highest harmonics observed have energies extending to >450eV [2], but the very low effective nonlinearity of the generating medium employed, He, coupled with the absence phase matching, limits the HHG signal to several hundred photons/harmonic peak/pulse. As a result, the most significant need for further advances remains in increasing the overall conversion efficiencies and pulse energies that can be achieved using high harmonic generation (HHG).

In general, the length over which the HHG signal will add constructively, known as the coherence length ($L_c$), is set by the wavevector mismatch, $\Delta k$, between the harmonic and driving laser E-fields, and is defined as

$$L_c = \frac{\pi}{|\Delta k|}, \quad \Delta k = k_q - qk_0 \quad (1)$$

with $k_q$ the wavevector of the $q^{th}$ harmonic order and $k_0$ the laser wavevector.

For typical high intensity HHG experiments in which an ultrashort laser pulse is focused into a gas jet, there are several factors affecting $\Delta k$: geometric ($\Delta k_g$, net negative contribution); free electron, or ionisation, induced ($\Delta k_e$, net negative contribution); and neutral gas density contributions ($\Delta k_a$, net positive contribution) [4, 5]. $\Delta k_g$, due mainly to the Gouy shift, can be replaced by the more controllable dispersion of hollow gas filled capillaries [4, 5, 6]. The use of capillaries also allows high intensity conditions to be maintained over extended interaction lengths with near constant/controllable pressure. In general for HHG experiments, the contribution of $\Delta k_e$ and $\Delta k_g$ to the wavevector mismatch in the nonlinear medium dominates $\Delta k$ under conditions of high average-ionisation ($Z^* = N_e/N_a \geq \sim 1$, where $N_e$ and $N_a$ are the number density of electrons and atoms respectively), and can be approximated simply as

$$\Delta k \approx \Delta k_e + \Delta k_g \quad (2)$$

In close analogy to the generation of optical harmonics of laser radiation in nonlinear crystals, high efficiency can be achieved by phase-matching the harmonic generation process throughout the non-linear medium i.e. setting $\Delta k$ to 0. Phase matching of low energy (<50eV) HHG from capillaries has been demonstrated with dramatic improvements in the observed signal obtained by pressure tuning to balance the free electron, neutral gas density and capillary propagation contributions to $\Delta k$ around specific atomic resonances [4]. However this scheme is limited to low levels of ionisation and hence low harmonic orders.

An alternative method for increasing the observed HHG signal is quasi-phasematching (QPM). In QPM no attempt is made to achieve $\Delta k = 0$, instead harmonic production is suppressed for half of the QPM period

$$L = 2mL_c \quad (3)$$

to prevent destructive interference between HHG from different regions of the medium, where $m$ (odd only) indicates the order of the QPM process. In analogy to the phasematching condition given in Eq. 1, one can write for quasi-phasematching

$$\Delta K = mK - |\Delta k| \approx 0 \text{ where } K = \frac{2\pi}{L} \quad (4)$$

where $\Delta K$ is the mismatch relevant to the QPM process.

Pioneering work has demonstrated the validity of the QPM scheme [5, 7, 8]. Paul, *et al.* [5] and Gibson, *et al.* [8], both using capillaries with a modulated bore diameter to vary the intensity of the driving laser in the capillary, show that enhancements in harmonic signal can indeed be achieved with some success. Paul, *et al*. [5] demonstrated enhancement by factor of order ~10 in the plateau region, while Gibson, *et al*. [8] demonstrated similar enhancements in the cut off regions of HHG spectra [8]. However, extending QPM to higher photon energies is difficult for current QPM schemes since $L_c$ is then typically <50$\mu$m. Moving to higher order *m* (Eq. 4) does allow signal enhancement at short wavelengths, but at significantly reduced signal growth rates.

In this work a new method of QPM enhancement of the high harmonic signal generated in hollow gas filled capillaries is presented, with substantial improvements in the HHG signal extending to high photon energies. This QPM process takes advantage of mode beating [9, 10, 11, 12] in capillaries (multimode quasi-phase matching, or *MMQPM*), to achieve very significant periodic intensity modulations over the entire interaction length. Excitation of these higher order capillary modes, and their subsequent beating, forms significant spikes in the axial intensity, separated by extended regions of low ionisation (>$10^{-2}$ of peak), allowing the HHG signal to grow via QPM. This process is shown to allow the generation of bright soft x-ray harmonic radiation with enhancements of >$10^4$ over the signal expected from a single coherence length and, as a consequence of the deep intensity modulations, to increase the quasi phase-matching period, L, at the highest harmonic orders.

## 2. Multimode quasi-phase matching

When a free space propagating laser beam is coupled into a hollow capillary waveguide a number of capillary eigenmodes are excited. For an axially centred, radially symmetric beam, azimuthal modes, $EH_{1,j}$, are dominantly excited, such that

$$E_L(r) = \sum_{j=1}^{\infty} C_j E_{1,j}(r) \tag{5}$$

where $E_L(r)$ and $E_{1,j}(r)$ are the transverse, or radial, electric field (E-field) distributions of the incident laser beam and $j^{th}$ eigenmode respectively. The relative peak on-axis E-field strength coupled into the $j^{th}$ mode, is given by the coupling coefficient $C_j$ [11] and the incident pulse intensity profile, $A(r) = |E_L(r)|^2$, in terms of the individual eigenmode profiles, is given as

$$A(r) = \left| \sum_{j=1}^{\infty} J_0\left(\frac{r}{a} \cdot U_{0,j}\right) C_j \right|^2 \tag{6}$$

where $U_{0,j}$ is the $j^{th}$ root of the zero order Bessel function, $J_0$, *a* is the capillary radius and *r* is the radius of the profile. Best coupling (~98% incident energy) of a Gaussian $TEM_{00}$ free space mode into the fundamental mode, $EH_{11}$, of the hollow capillary is achieved for

$$\chi = \frac{w_0}{a} = 0.64 \tag{7}$$

where $\chi$ is referred to as the coupling parameter, $w_0$ is the $1/e^2$ radius of the incident beam, and *a* is the capillary radius [9]. For $\chi \neq 0.64$ more energy is coupled into higher order modes. The fraction of the incident E-field coupled into each mode is determined by $\chi$ [9, 10], [11]. At a distance z into the capillary the axial intensity of the guided beam is given by

$$I(z) = \left| \sum_{j=1}^{\infty} C_j E_{1,j} e^{i(\Delta k_j \cdot z)} \cdot L_{z,j} \right|^2 \tag{8}$$

given that $L_{z,j} = Exp(z/z_j^{at})$, where $z_j^{at} (\propto a^3)$ is the attenuation length for the $j^{th}$ mode [10], and $\Delta k_j = \beta_1 - \beta_j$ and $\beta_j$, the capillary propagation wavevector of the $j^{th}$ mode, is given by

$$\beta_j = k_0 \left[ 1 - \frac{1}{2} \left( \frac{U_{0,j}}{k_0 a} \right)^2 \right] \quad (9)$$

where $k_o$ is the wavevector of the incident laser light [10].

## 3. Results and discussion

The characteristic high frequency beat length, $L_B$, of the intensity modulation, at a given point $z$ in the capillary is set by the highest order mode, $jmax$, with a significant amplitude after propagating a distance equal to $z_{jmax}^{at}$ [10]. Importantly, in the *MMQPM* schemes presented here, $L_B$ at a given point $z$ in the axial intensity profile, $I(z)$, is not set by *the number of modes*, $j^*$, included in the truncated calculation i.e. $L_B$ is not a function of $j^*$.

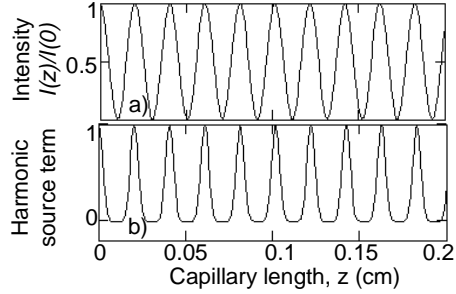

Fig. 1. Idealised multimode beating assuming two equal intensity modes ($j$=1 and $j$=20) excited in a capillary with $a$~90$\mu$m. Fig. 1a) shows the axial intensity as a function of z, $I(z)$, normalised at $z$=0, over the first 0.2 cm of the capillary while Fig. 1b) shows corresponding harmonic source term (assumed proportional to the ADK rate for tunnel ionisation [13]).

In an ideal QPM scenario one would have perfectly periodic intensity modulations and this could be achieved for two excited modes with equal intensity (Eq. 8). This idealised case is shown in Figure 1 for $jmax$=20 (giving $L_B$~200$u$m), and assuming $z_{jmax}^{at} \to \infty$, in an evacuated capillary with $a$~90$\mu$m.

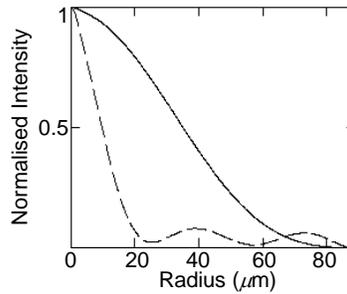

Fig. 2. Normalised Intensity profiles for coupling to an evacuated capillary with $a$=90$\mu$m. The coupling coefficient for the Gaussian (solid line) and the central maximum of the Airy profile (dashed line) are $\chi$=0.64 and $\chi$=0.2 respectively.

For realistic transverse profiles of the beam incident on the waveguide, a large number of waveguide modes – as opposed to two – will be excited. In this section we investigate *MMQPM* for the two possible transverse intensity profiles presented in Figure 2: that of an ideal Gaussian beam profile and, more in line with that typical of many high power lasers with flat-top/super Gaussian near-field distributions, an Airy profile.

The on axis intensity profiles (Eq. 8) in an evacuated capillary with $a$=90$\mu$m for values of $C_j$ (Eq. 5) resulting from the Gaussian ($\chi=0.64$) and Airy ($\chi=0.2$ for the central maximum) profiles shown in Figure 2 are presented in Figures 3a) and 3b) respectively.

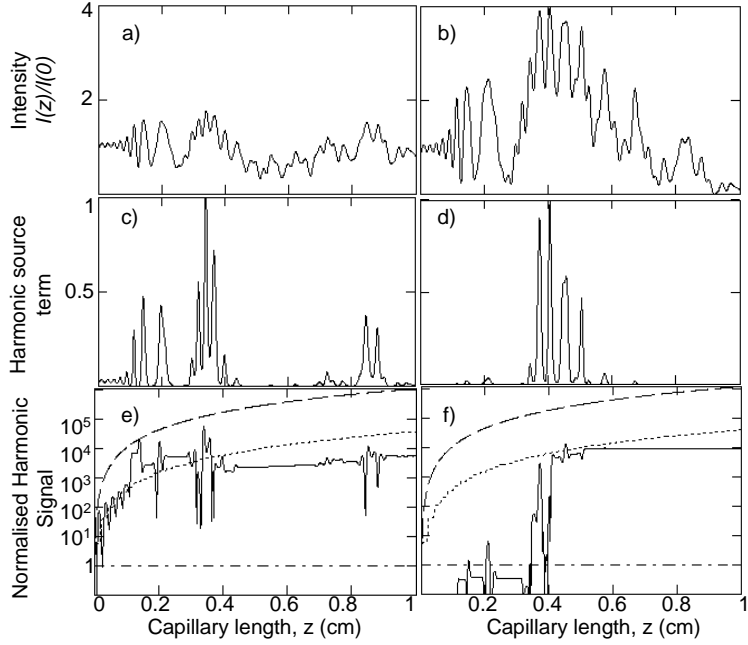

Fig. 3. Multimode quasi phase matching. Calculated axial intensity as a function of distance $z$ into a capillary of radius 90$\mu$m for beams with the Gaussian (a) and Airy (b) incident transverse profiles shown in Fig. 2. The corresponding calculated harmonic source terms (normalised at peak intensity) are shown in Fig. 3c) and 3d). The resulting harmonic growth for q=201 (solid lines), normalised to the signal expected for one coherence length (dot-dashed line), is shown in Fig. 3 e) and f). The dotted line in both 3e) and 3f) corresponds to the expected harmonic growth for the two mode axial intensity profile given in Fig. 1a), while the dashed line is the signal expected for perfect phase matching. Harmonic growth is calculated using a 1-D code under optimal matching conditions for the normalised harmonic source terms given in Fig. 3c) and 3d). Optimal matching conditions were achieved in each case by varying intensity at $z$=0 and pressure (approx. 20mbar Ar) slightly to adjust $Z^*$ such that $2L_c \approx L_B$ over extended interaction regions. It should be noted that the rapid growth of HHG over the first few mm of the capillary for the intensity profiles in Figures 3c) and 3d) in comparison to that for the idealised *MMQPM* scheme from Figure 2 is due to the constructive interference of multiple modes leading to increased on-axis intensity and hence greater HHG signal.

As can be seen, even for the optimal coupling of a transverse Gaussian beam profile to the fundamental mode, $j$=1, (Eq. 7), significant intensity modulations (>50%) can exist in the capillary over $z_j^{at}$ [10] of the higher order modes. The Airy profile couples more energy into higher order modes leading to increased on-axis beating, as can be seen in Figure 3b). The transverse intensity profile in Figure 2 (dashed line) approaches that expected for perfect matching to eigenmode $j$=3 (Eq. 6) giving enhanced low frequency modulation in the corresponding axial intensity profile while the high frequency modulation ($L_B$) is still set by the highest order significantly contributing mode. These profiles provide the platform for a significantly modulated on-axis harmonic source term as shown in Figures 3c) and 3d) respectively.

The existence of sharp peaks in intensity, spaced by multiples of $L_B$, can lead to quasi-phasematching and consequently rapid growth in harmonic signal, as shown in Figures 3e) and 3f) (solid line) for q=201. The signal growth is calculated using a 1-D code assuming 40fs pulse and ~20mbar Ar present in the capillary (the signal for q=201 is from $Ar^+$) where the harmonic source term is assumed to be proportional to the ADK tunnelling ionisation rate [1, 13]. From the results presented in Figure 3, the concept of *MMQPM* is clearly quite robust with regards to incident transverse intensity profiles i.e. as long as the intensity and pressure are chosen correctly it is possible to achieve substantial signal growth for a variety of input profiles. Low harmonic order *MMQPM* (low $\Delta k$) can be readily achieved by tuning pressure and intensity accordingly.

As can be seen (in this example) most of the growth for high $\Delta k$ takes place over the first few mm of the capillary before the higher order modes are significantly damped and hence it might appear that *MMQPM* may not be readily scaled to longer lengths. However, it should be noted that $I(z)$ has been calculated assuming evacuated capillaries (i.e. no gas present). It is expected that the presence of gas in the capillary will cause continuous repartition of energy to higher order modes [10] thereby extending the distance over which *MMQPM* operates. Furthermore, it may also be possible to force mode mixing suitable for *MMQPM* by slightly bending, or changing the bore radius of, the capillary.

We have recently demonstrated dramatic enhancements ($>10^3$, conversion efficiencies $>10^{-6}$) at soft x-ray wavelengths (extending to 360eV, q=231) generated by high harmonic generation driven within an argon-filled straight bore capillary waveguide. This result, which we interpret as arising from *MMQPM*, will be presented in an imminent publication elsewhere [14].

Under conditions for which Eq. 4 can be satisfied, shown in Fig. 3 (c) and (d), the rapid variations of $I(z)$ correspond to imperfect $m$=1 QPM processes i.e. all the destructive contributions are '*switched off*' but not all the constructive contributions are '*switched on*'. As a result of the strong intensity modulation, the degree of ionisation, $Z^*$, is high at points of strong HHG and low everywhere else. This results in a much lower average $Z^*$, and hence lower $\Delta k$, allowing a longer – and, critically, easier to achieve – QPM period to be employed (from Eq. 4). This is particularly important for QPM of very high-order harmonics, for example those beyond the C K-edge (~4.3nm, harmonic order, q, ~ 185 for ~800nm fundamental). In essence, *MMQPM* allows much longer $L_c$ in comparison to a flat intensity case.

An important corollary of this work, in relation to HHG, is that even under optimal coupling conditions significant modulation of the on axis intensity can occur when HHG is driven within a waveguide. If these are to be avoided, such as when attempting true phase-matching, harmonic generation should only be undertaken at sufficiently long distances into the waveguide for all but the lowest-order mode to have been absorbed.

**4. Conclusion**

In conclusion, we have proposed a new scheme for quasi phase-matching which utilizes fluctuations in the axial intensity of a beam propagating in many modes of waveguide Numerical modeling shows that this approach can enhance the intensity of harmonics by a factor of order $>10^4$ even for harmonics beyond the C K-edge, for which the coherence length can be very short. Given the inverse dependence of $C_j$ and $\beta_j$ with respect to capillary bore radius, $a$, the mode structure in the capillary therefore depends strongly on $a$. Our calculations indicate that moving to smaller capillary bore diameters will allow the generation of higher frequency mode beating profiles suitable for QPM at very high $\Delta k$.

MZ is grateful to the Royal Society for support through a Wolfson Merit Award, SMH to the Engineering and Physical Sciences Research Council for research funding (EP/C005449), and ML to the Rhodes Trust for a scholarship.